# Relative Condition Factors of Fish as Bioindicators One Year after the *Deepwater Horizon* Oil Spill

Joshua Courtney,[1] Taylor Klinkmann,[2] Amy Courtney,[1] Joseph Toraño,[2] and Michael Courtney[2]

[1]BTG Research, P.O. Box 62541, Colorado Springs, CO, 80962
[2]United States Air Force Academy, 2354 Fairchild Drive, USAF Academy, CO, 80840
Michael_Courtney@alum.mit.edu

**Abstract**
Creel surveys were performed over a three week period in late spring, 2011, in the Lafourche and Calcasieu area estuaries of the Louisiana Gulf Coast. Weights and lengths were measured for black drum (*Pogonias cromis*), red drum (*Sciaenops ocellatus*), and spotted seatrout (*Cynoscion nebulosus*), and relative condition factors were calculated relative to expected weights from the long term (5 year) Louisiana data for each species. A normal relative condition factor is 1.00. The mean relative condition factors in the Lafourche area were black drum, 0.955 ± 0.020; red drum, 0.955 ± 0.011; spotted seatrout, 0.994 ± 0.009. In the Calcasieu area, the mean relative condition factors were black drum, 0.934 ± 0.017; red drum, 0.965 ± 0.014; spotted seatrout, 0.971 ± 0.010. Results suggest that the abundance of primary food sources for black drum and red drum in Lafourche, including oysters and crab, were likely affected by the oil spill and continued to be reduced one year later. Increased harvest of oysters and blue crab in the Calcasieu area (in part to make up for the ban in most of Louisiana) resulted in less food for the black drum and red drum there, also. Spotted seatrout eat mainly shrimp and small fish and showed no significant reduction in relative condition factor in Lafourche and a slight reduction in Calcasieu one year after the oil spill.

**Introduction**

It is not known how long the *Deepwater Horizon* Oil Spill in the Gulf of Mexico in 2010 will affect the health of the ecosystem.  Large areas were closed to fishing, but others were opened to oyster harvest for the first time in several years.  It is difficult to measure the quantity of forage species, though these are important both for the health of the ecosystem and for human consumption. By using predator species as bioindicators, an estimate can be made of how the ecosystem was affected by the oil spill.

Bioindicators are biological features that can be measured and that could change with exposure to negative environmental factors (Summers et al. 1997; Bortone 2003). Sometimes only part of an organism is affected by a given negative factor. Condition indices are often used in fisheries and some are based on weight and length of the animal. A morphological condition index is a whole animal bioindicator because it may be affected by many factors (Barton et al. 2002).  A fish that is heavier for a given length (higher condition index) is generally considered to be a healthier fish, because extra weight means extra energy reserves. Lighter fish lack energy reserves and tend to be more susceptible to environmental stressors.  A low body condition may also suggest muscle wasting (proteolysis) indicating a starvation response (Barton et al. 2002).  It has also been suggested that females with a lower body condition reduce reproductive investment yet still have an increased risk of mortality (Lambert and Dutil 2000).

Two areas were selected for this study, estuaries near bayou Lafourche on the central Louisiana coast and the Calcasieu estuary in southwestern Louisiana.  The Lafourche area was chosen because it has active sport and commercial fisheries which were closed for most of 2010 after the oil spill and is among the shore areas where the oil spill made landfall and had a large effect on wildlife.  The Calcasieu estuary was chosen because it also has active sport and commercial fisheries.  It is over 100 km west of the areas of fishing closures or areas known to be directly impacted by the oil spill.  Consequently, impact of the oil spill in the Calcasieu estuary would more likely be due to changes in fishing pressure as sport and commercial fisherman concentrated efforts in coastal areas of Louisiana that remained open to fishing through all of 2010.

Three species of fish – black drum (*Pogonias cromis*), red drum (*Sciaenops ocellatus*), and spotted seatrout (*Cynoscion nebulosus*) - were selected for study. These spend most of their time in different parts of the water column, and they have different but somewhat overlapping feeding habits. Black drum eat at the seafloor and prefer shellfish like oysters and bottom-dwelling crustaceans like crab (Matlock 1990). Red drum feed throughout the water column and prefer to eat shrimp, smaller fish, and a lot of crab (Pearson 1929).  Spotted seatrout feed mostly near the top of the water column and prefer eating shrimp and very small fish (Blanchet 2001). Finally, these three species are common to the Louisiana Gulf region and are popular catches for fishermen. Therefore, many fish were available to measure in creel surveys.

*Black Drum Distribution and Feeding*

Black drum are commonly found throughout the Gulf Coast including the Calcasieu estuary and the Lafourche area estuaries.  Black drum spawn in estuaries in February and March (Sutter et al. 1986), then are usually found near the bottom of the water column in partially muddy water (Jenkins 2004). Silt from rivers or wave action are the cause of the opaque water. Their chin barbels help them locate food in this environment. Larger black drum prefer the saltier water closer to the ocean.

About one-third of the diet of smaller adult black drum is mollusks, while the rest of the diet is made of small fish, invertebrates and shrimp. Larger adult black drum are mostly bottom feeders, and up to two-thirds of their diet consists of oysters, when available (Sutter et al. 1986). Most of the rest of their diet is crabs and shrimp, with studies showing some differences with location and availability. Black drum spend a lot of time over oyster beds.

Studies have shown that adult black drum can eat two regular sized oysters or crab per kilogram of their body weight per day (Cave and Cake 1980). Black drum are known to destroy large quantities of oysters on seed reefs and lease areas in Mississippi and Louisiana (Benson 1982) and there are active research programs to determine methods of reducing this damage (Brown et al. 2006; Brown et al. 2008).

*Red Drum Distribution and Feeding*

Red drum are also commonly found in the Calcasieu estuary and the Lafourche area estuaries. Red drum naturally occur along the southern Atlantic and Gulf of Mexico coasts of the United States, including the coasts of Louisiana, Alabama, Mississippi, and Florida. Immature red drum prefer grass marsh areas of bays and estuaries when available. Both younger red drum (3-6 years of age) and bull red drum prefer rocky outcroppings including jetties and manmade structures, such as oil rigs and bridge posts. Around this type of structure, they are found throughout the water column (Matlock 1990). Some scientists have observed gradual movement between shallower and deeper areas with the seasons, preferring estuarine environments during summer (Jenkins 2004). In an early study Pearson (1929) reported that in the winter, young red drum (5-15 cm long) off the Texas Gulf coast would move from shallow coves into deeper bayous, and that larger, young red drum (40-60 cm long) would gradually move from the bayous into deeper bays, returning to the shallower places in the spring. Pearson observed that red drum longer than about 70 cm traveled in schools and were found along the sandy shores as well as the bays of the Gulf of Mexico.

Red drum start to eat their characteristic foods of shrimp, fish and crabs after they are longer than 100 mm which typically occurs six to seven months after hatching (Peters and McMichael 1987). The red drum uses its senses of sight and touch and its down turned mouth, to locate forage on the bottom through vacuuming or biting the bottom. On the top and middle of the water column, it uses changes in the light to identify things that look like food. Their diet changes some with the seasons due to the relative amount of different foods. They feed on crabs all year; in the spring they have been found to eat more shrimp; in the fall they have been found to eat more menhaden and small bait fish (Scharf and Schlight 2000).

*Spotted Seatrout Distribution and Feeding*

Spotted seatrout, also called speckled trout, are commonly found in the Calcasieu estuary and the Lafourche area estuaries. In general, they are numerous throughout the Gulf of Mexico, and they are also found along the Atlantic coast. They spawn in April through July and generally prefer estuarine environments (Jenkins 2004). They range from deep inland in estuaries into the Gulf up to waters 30 feet deep, though they prefer marshy areas with sea grasses. Studies in several Gulf states showed that most spotted seatrout stay close to one bay system throughout their lives. They swim in schools and move inland to less salty water in the winter and back out to saltier water in the summer months (Johnson and Seaman 1982).

In general, the spotted seatrout population seems to be thriving in Louisiana. In fact, Louisiana named it the official state saltwater fish in 2001. Also, the Department of Wildlife and Fisheries has made generous catch limits that were 25 speckled trout per person per day at the time of the study (with a few restrictions).

Spotted seatrout are active fish and eat a lot, especially during the summer. Fish under 500 – 600 mm (up to 2 years old) eat mostly shrimp (*penaeidae*) and crustaceans along with a variety of other foods. As they grow larger, their preferences shift to fish. At first, they prefer smaller fish such as silversides and anchovies. As the spotted seatrout continues to grow, it moves to larger prey fish such as menhaden, croakers and mullets. The largest seatrout (above 600 mm) have a strong preference for the largest mullet they can physically eat (Horst 2003). There are much fewer of these biggest spotted seatrout, and they tend to be more solitary, female, and over 6 years old.

**Methods**

Creel surveys were performed over a three week period from late May to mid-June, 2011. In each of two study areas, anglers were asked for permission to weigh and measure fish from their creels. For reasons described above, the species studied were black drum, red drum, and spotted seatrout.

One study area was the Calcasieu estuary in southwestern Louisiana centered near +29° 56' 45.64 N", -93° 18' 12.84" W and extending approximately 20 km north and south and 10 km east and west from this point, including 2 km into the Gulf of Mexico near Calcasieu pass. Specifically, creel surveys were performed at boat launches at Calcasieu Point, Hebert's Marina, and the Cameron Jetties (referred to as Calcasieu from this point). The other study area was in Lafourche Parish, Louisiana, centered near +29° 15' 10.28"N, -90° 12' 44.24" W and extending approximately 25 km north and south and 15 km east and west and included bayou Lafourche from Golden Meadow, south to the estuaries including Barataria Bay and Timbalier Bay and up to 2 km south of the Gulf Coast (referred to as the Lafourche area from this point). Specifically, creel surveys were performed at Bobby Lynn's Marina and the Port Fourchon public boat launch. The reason for measuring fish from two geographic areas was that the two areas were different distances from the source of the oil spill and fishing closures were different in each area; therefore, the environment might have been affected differently.

Each fish was weighed to the nearest 0.01 kg. The fork length (length from the mouth to the vertex of the fork in the tail) and total length (length from the mouth to the longest point of the tail) of each fish were measured to the nearest 3.2 mm (1/8 inch). A total of 43 black drum (29 from Calcasieu and 14 from Lafourche), 141 red drum (66 from Calcasieu and 75 from Lafourche), and 332 spotted seatrout (138 from Calcasieu and 194 from Lafourche) were measured.  Using published weight-length equations for each species for Louisiana waters (Jenkins 2004), a relative condition factor was computed for each fish that was measured.  The expected weight equations from Jenkins (2004)

Weight-length equations for fish have the form, $W = aL^b$, where *W* is the weight (g), *L* is the total length (mm), and *a* and *b* are values found by fitting to measurement data for a given species of fish. The power *b* is close to 3 for most species of fish.

Using weight and length data, there are at least three ways to express an overall condition index (Stevenson and Woods 2006).  One is the Fulton condition index, K, in which the power *b* is defined to be exactly equal to 3, and K is equal to the actual weight of a fish divided by its total length raised to the third power. This method, though common, is not the best choice when a wide range of lengths is included in a sample. Another common way to express the condition of a fish is through the relative weight, which is the actual weight of the fish divided by its expected weight based on a standard weight-length equation.  Standard weight equations exist for many species of freshwater fish but not for the saltwater species measured in the present study.  Saltwater species tend to have a much wider geographic range, and weight-length relationships developed for the species measured in the present study have shown geographic differences.

A third common way to express condition index for a fish was used in the present study.  It is termed the relative condition factor, Kn (LeCren 1951; Sutton et al. 2000), and is computed as $Kn=W/aL^b$, where W is measured body weight of the fish in question (g), L is total length (mm), and a and b are the parameters from an appropriate reference weight-length equation.  A relative condition factor greater than one means the fish weighs more than expected for its length; a relative condition factor less than one means the fish weighs less than expected for its length.

Jenkins (2004) reported both spring and fall weight-length relationships from a long term (5 year)  state wide data set.  The spring relationships were used here for the reference weight length equations, since these data were from the same season as the samples in the present study.  For black drum and spotted seatrout, separate weight-length relationships were published for male and female fish.  In this study, the sex of each fish was not recorded.  Therefore, the spring relationships for these species were used to develop appropriate combined weight-length relationships.

For black drum and spotted seatrout, length values from 0 to 1000 mm and 0 to 600 mm, respectively, in increments of 10 mm, were entered into a spreadsheet program. The spring weight-length equation for males for each species was used to calculate corresponding length values.  In a similar way, the spring weight-length equation for females for each species was used to generate a set of length-weight data points.  The data points for male and female fish were combined, and a weight-length relationship was then fit to the combined data set. The resulting equations (Table 1) for black drum and for spotted seatrout were used to compute relative condition factors in the present study.

*Table 1 Reference weight-length equations based on Spring relationships published by Jenkins (2004). Weight is in grams and length is in millimeters.*

| Species | Sex of Fish | Weight-Length Equation |
|---|---|---|
| Red Drum | **Combined** | **W = 0.000005297L$^{3.110}$** |
| Black Drum | Male | W = 0.000005821L$^{3.149}$ |
| | Female | W = 0.000008241L$^{3.093}$ |
| | **Combined** | **W = 0.000006926L$^{3.121}$** |
| Spotted Seatrout | Male | W = 0.000011535L$^{2.969}$ |
| | Female | W = 0.000007834L$^{3.035}$ |
| | **Combined** | **W = 0.000009506L$^{3.00}$** |

**Results**

A summary of the results is shown in Table 2. Graphs of relative weight vs. total length are shown in Fig. 1. The uncertainty in each mean was computed as the standard error of the mean (SEM). The total lengths measured covered the range of mature lengths for these species. Some observations can be made by comparing the condition factors measured for each species of fish and at each location (Fig. 2). All of the fish had mean (average) relative condition factors that were lower than 1.00. However, if the uncertainty in the mean is considered, the relative condition factor for the spotted seatrout from Lafourche (at 0.994 ± 0.009) is not significantly different from 1.00. The relative condition factor for spotted seatrout from Calcasieu was below normal, at 0.971 ± 0.010.

*Table 2 Summary of results for black drum, red drum, and spotted seatrout for samples measured May-June, 2011 in the Lafourche and Calcasieu areas of the Louisiana Gulf Coast. The uncertainty for each mean was computed as the standard error of the mean.*

| | Sample Size | Total Length Range (mm) | Weight Range (g) | Mean condition factor | Uncertainty | p-value |
|---|---|---|---|---|---|---|
| Black Drum | | | | | | |
| Lafourche | 13 | 673-965 | 3877 - 14371 | 0.955 | 0.020 | 0.027 |
| Calcasieu | 28 | 308-692 | 332 - 4290 | 0.934 | 0.017 | < 0.001 |
| Red Drum | | | | | | |
| Lafourche | 75 | 400-1048 | 665 - 12648 | 0.955 | 0.011 | < 0.001 |
| Calcasieu | 66 | 406-1013 | 624 - 10312 | 0.965 | 0.014 | 0.012 |
| Spotted seatrout | | | | | | |
| Lafourche | 194 | 279-527 | 211 - 1289 | 0.994 | 0.009 | 0.476 |
| Calcasieu | 138 | 292-571 | 191 - 1672 | 0.971 | 0.010 | 0.003 |

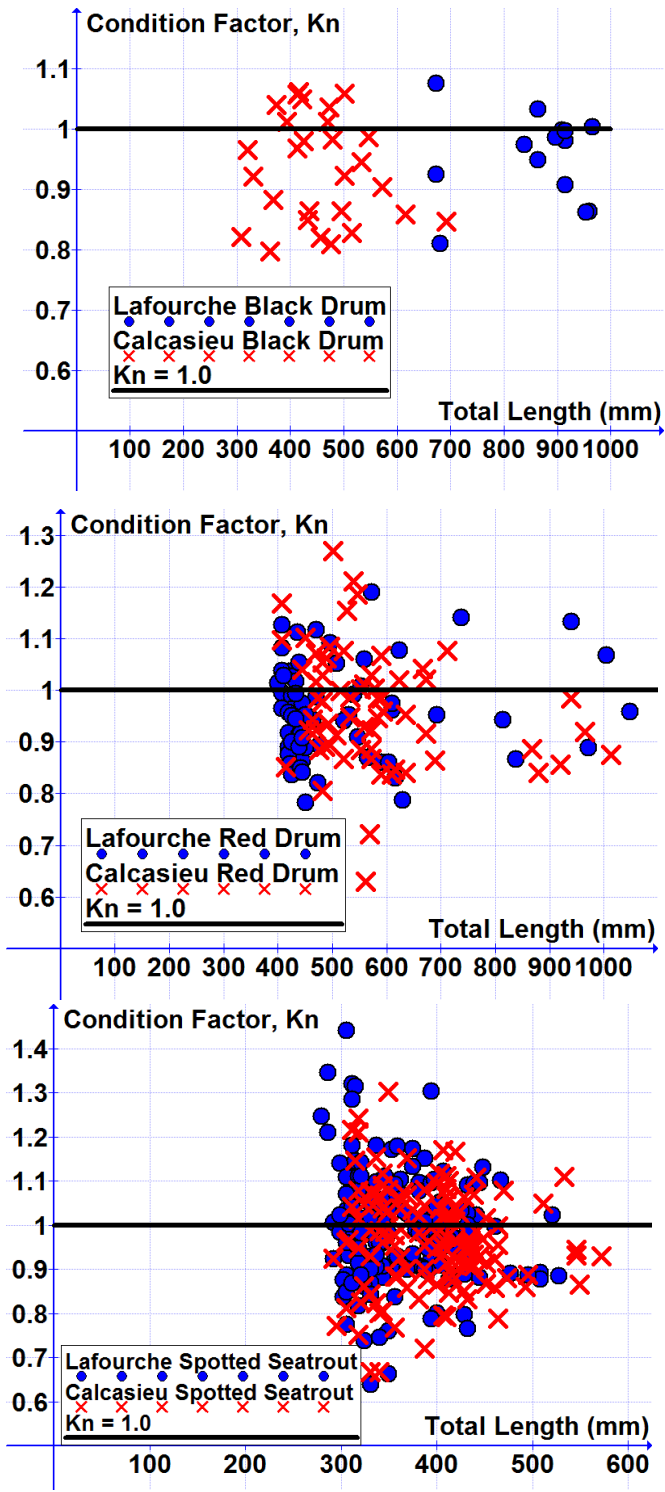

**Fig. 1** Relative condition factor vs. total length for black drum (top), red drum (middle), and spotted seatrout (bottom)

Fig. 1 (top) shows the relative condition factor vs. length for back drum. The specimens available in the Lafourche area estuaries tended to be longer than those in the Calcasieu estuary. Most of the samples from both locations tended to be below the expected weight for Louisiana waters. Fig. 1 (middle) shows the relative condition factor vs. length for red drum. All six of the largest specimens (over 800 mm) for Calcasieu have condition indices below 1.00, but that only four of the six largest specimens in Lafourche have relative condition factors below 1.00. Fig.1 (bottom) shows the relative condition factor vs. length for spotted seatrout, which appears to be more evenly distributed about the value 1.00 than in black drum or red drum.

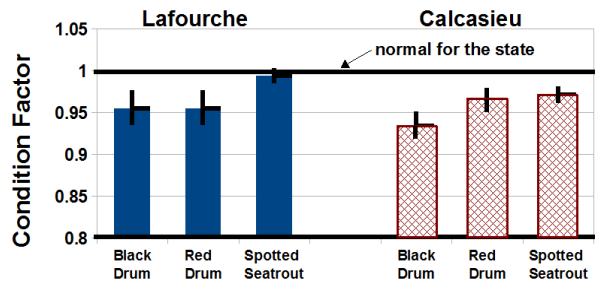

*Fig. 2 Summary of results for mean relative condition factors (and uncertainties in the means) in Lafourche and Calcasieu regions of the Gulf Coast of Louisiana in late May and June, 2011*

Fig. 2 summarizes the results of the mean relative condition factor for each of the three species at each of the two study areas. In the Lafourche area black drum, red drum, and spotted seatrout had mean condition factors of 0.955 ± 0.020, 0.955 ± 0.011, and 0.994 ± 0.009, respectively. Smaller uncertainties reflect the effect of larger sample sizes in reducing uncertainty in the mean rather than a narrower distribution. In the Calcasieu Estuary black drum, red drum, and spotted seatrout had mean condition factors of 0.934 ± 0.017, 0.965 ± 0.014, and 0.971 ± 0.010, respectively.

**Discussion**

As a whole animal bioindicator (Barton et al. 2002) there may be situations where reductions in condition factor reflect environmental stressors other than reduced forage availability, and ruling out other factors is challenging in any given case. In the present study, ruling out the direct effects of oil or dispersants is not possible for samples taken in the Lafourche study area. Nevertheless, in most cases reduced availability of forage is probably a contributing factor to significant reductions in body condition.

The relative condition factors of black drum and red drum were significantly lower than 1.00 in the Lafourche area estuaries. Oysters, which make up a significant part of the food supply of black drum are well known to have been significantly impacted by the oil spill in this area, and it is likely that other mollusks that black drum depend on for food were affected as well. There have been news reports and complaints from crab fishermen about reductions in blue crab harvest by 70-75%, but on the whole reduced availability of these species has not yet been as well documented as the reduction in oyster availability.  Anecdotally, the study team did observe many crab traps stacked in disuse on roadsides and in parking lots, and it did seem that crabs were more difficult than usual to obtain from local vendors who complained that the crab population was not making it worthwhile to maintain operational traps south of Leeville.  If other decapods with similar life histories to the blue crab were similarly impacted by the oil spill, this might account for the reduced body condition of black drum and red drum in the Lafourche area estuaries.    An earlier study after a much smaller (600 barrels) Barataria Bay oil spill (Roth and Baltz 2009) showed that the population and structure of the benthic decapod community seemed to recover quickly.

In a 2004 study including body condition of contaminated and non-contaminated sites in the Calcasieu estuary, Jenkins (2004) found a condition factor of 1.03 ± 0.02 at the non-contaminated sites and 0.95 ± 0.01 at the contaminated sites using the same reference data but combining species to obtain sample sizes of n=23 at the reference sites and n=122 at the contaminated sites.  This suggests that the species measured in the present study are fairly sensitive to environmental stressors, and that in the absence of any stressors, the relative condition factor of these species in the Calcasieu estuary would be expected to be closer to Louisiana as a whole than the findings of the present study.  Given that neither the oil nor the dispersants reached the Calcasieu estuary, the observed body condition significantly below 1.00 is most likely due to another cause.

The federal government and state of Louisiana closed the areas around Bayou Lafourche to fishing for months after the oil spill.  There were two reasons for this. One reason was to protect people from eating seafood that might be unsafe. The second reason was to protect the species that suffered losses due to the oil spill, so that they could reproduce and return to normal population size faster. To help the oyster fishermen make up for the closures in the Lafourche area, Louisiana opened oyster fishing in Calcasieu, which had not been opened to oyster fishing for several years. Louisiana closed some of the areas one month before the planned closing date because too many oysters were being taken (Louisiana Department of Wildlife and Fisheries 2011).

The fishing management decisions help to explain the results of this study. The food supply for the black drum in Lafourche was most likely affected by the oil spill killing or poisoning forage species including mollusks and crustaceans. The food supply for the black drum in Calcasieu was probably affected by the supply of oysters and crab being decreased by increased harvest pressure.  Since the red drum also eat a lot of crab and other decapods that might have been killed as bycatch of increased shrimping pressure, the decreased relative condition factor measured in the present study may have similar causes. However, spotted seatrout eat mostly shrimp and small fish. Since the relative condition factor for spotted seatrout was close to normal one year after the oil spill, it seems that the shrimp and small fish species might not have experienced as significant effects from the oil spill or might have recovered more quickly.  It is ironic that increased harvest pressure in an area that remained open, caused in part by large areas of closure, seems to have had at least a large an impact on body condition of the three indicator species a year later than direct effects of the oil spill itself.

Even though the relative condition factor of spotted seatrout in Lafourche did not indicate significant reduction in body condition, it is still possible that growth was reduced or that more sensitive bioindicators will reveal effects of the oil.  A research group from the University of Southern Mississippi is also studying spotted seatrout and red drum as bioindicators of the effects of the Deepwater Horizon oil spill using more sensitive techniques (Brewton et al. 2011; Fulford et al. 2011).  The researchers are looking at body size, molecular biomarkers, histology, and growth rings in otoliths. The study is focusing on whether exposure to oil affected the health, growth, and ability to reproduce of these species. The studies are scheduled to be finished by the spring of 2012.

Other possible reasons for reduced body condition should be discussed.  Higher stock sizes leading to more intra-specific competition is a possibility.  This possibility was not supported by the anecdotal observations of most anglers finding the fishing difficult and area charter boat captains reporting having to travel far and wide and having difficulty finding spotted seatrout and red drum.  Another possibility is increased inter-specific competition. Though no formal stock assessments were performed, there is nothing to suggest from either reports of either commercial or recreational fisheries to suggest that any local species of fish was particularly thriving so as to be strongly out competing other species.

It is possible that the lower body condition of breeding size red drum and black drum may impact reproductive success and productivity of these fisheries in future years.  Lambert and Dutil found that the decline in condition and energy reserves of cod during several consecutive years may have lowered the productivity of the Northern Gulf of St. Lawrence stock, which combined with overfishing, could have contributed to the collapse of this stock (Lambert and Dutil 1997).  In a study of captive cod, fecundity was found to be significantly lower in groups with lower body condition (Kjesbu et al. 1991).  The gonosomatic index of red drum is correlated with body condition (Overstreet 1983) but it is unclear if fecundity in red drum is greatly reduced by the reduction in body condition reported here.  Likewise, little definitive information is known about how the fecundity of black drum relates to body condition (Fitzhugh et al. 1993; Sutter 1986). Future studies will do well to note whether there are significantly fewer red drum and black drum from the classes of 2010 and 2011.

**Acknowledgments**
This work was supported by BTG Research (www.BTGResearch.org).  The authors are grateful to Matthew Courtney whose encouragement, hospitality, and advice were essential throughout the planning and execution phases.  We also thank Bobby Lynn for his warm hospitality throughout our stay in Leeville and the charter boat captains and sport fisherman putting in at his marina who cooperated with the study.  The cooperation of charter boat captains and sport fisherman in Calcasieu was also essential.  Rachel Brewton brought valuable references to our attention.  We also appreciate feedback from Lt Col Andy Gaydon (USAFA/DFMS) and an anonymous reviewer whose comments were incorporated to improve the manuscript.

## Literature cited

Barton, Bruce A., John D. Morgan, and M Mathilakath M. Vijayan. 2002. Physiological and condition-related indicators of environmental stress in fish. In Biological Indicators of Aquatic Ecosystem Stress, ed. S. Marshall Adams, 111–148. Maryland: American Fisheries Society.

Benson, Norman G. 1982. Life history requirements of selected finfish and shellfish in Mississippi Spound and adjacent areas. U.S. Fish and Wildlife Service Biological Services Program FWS/OBS-81/51.

Blanchet, Harry, Mark Van Hoose, Larry McEachron, Bob Muller, James Warren, Joe Gill, Terry Waldrop, Jerald Waller, Charles Adams, Robert B. Ditton, Dale Shively, and Steve VanderKooy (ed.). 2001. The spotted sea trout fishery of the Gulf of Mexico, United States: A regional management plan. Publication number 87, Gulf States Marine Fisheries Commission, P.O. Box 726, Ocean Springs, Mississippi 39566-0726.

Bortone Steven A., ed. 2003. Biology of the Spotted Seatrout. Boca Raton, FL:CRC Press LLC.

Brewton, Rachel A., Robert J. Griffitt, and Richard S. Fulford. 2011. Impacts of the Deepwater Horizon oil spill on the health and growth of estuarine fish and ecosystem functionality. 2011 Northern Gulf Institute Annual Conference, Mobile, AL, May 17-19, 2011.

Brown, Kenneth M., Gerald J. George, Gary W. Peterson, Bruce A. Thompson, and James H. Cowan, Jr. 2008. Oyster predation by black drum varies spatially and seasonally. Estuaries and Coasts 31: 597-604.

Brown, Kenneth M., Gary W. Peterson, Gerald J. George, and Michael Mcdonough. 2006. Acoustic deterrents do not reduce black drum predation on oysters. Journal of Shellfish Research 25(2): 537-541.

Cave, R. Neil, and Edwin W. Cake, Jr. 1980. Observations on the predation of oysters by the black drum *Pogonias cromis* (Linnaeus) (Sciaenidae). Proceedings of the National Shellfisheries Association 70(1): 121.

Fitzhugh, Gary R., Bruce A. Thompson, and Theron G. Snider III. 1993. Ovarian development, fecundity, and spawning frequency of black drum *Pogonias cromis* in Louisiana. Fishery Bulletin U.S. 91: 244-253.

Fulford, Richard S., Robert J. Griffitt, and Nancy Brown-Peterson. 2011. Impacts of the Deepwater Horizon oil spill on the health and growth of estuarine fish and ecosystem functionality. 11-BP_GRI-23 – Active Project, Northern Gulf Institute (01/01/2011 – 02/29/2012) http://www.northerngulfinstitute.org/research/abstract.php?pid=150 Accessed 18 January, 2012.

Horst, Jerald. 2003. Speckled trout facts. Louisiana State University and Sea Grant Louisiana, http://www.seagrantfish.lsu.edu/pdfs/factsheets/speckledtrout.pdf. Accessed 18 January 2012.

Jenkins, Jill A. 2004. Fish bioindicators of ecosystem condition at the Calcasieu estuary. National Wetlands Research Center, USGS Report 2004-1323.

Johnson, Darlene R., and William Seaman, Jr. 1982. Species Profiles: Life histories and environmental requirements of coastal fishes and invertebrates (South Florida): Spotted sea trout. US Fish and Wildlife Service, Biological Report 82(11.43), TR EL-82-4.

Kjesbu, Olav Sigurd, J. Klungsoyr, H. Kryvi, Peter R. Witthames, and M. Greer Walker. 1991. Fecundity, atresia, and egg size of captive Atlantic cod (Gadus morhua) in relation to proximate body composition. Canadian Journal of Fisheries and Aquatic Sciences 48: 2333-2343.

Lambert, Yvan, and Jean-Denis Dutil. 1997. Can simple condition indices be used to monitor and quantify seasonal changes in the energy reserves of Atlantic cod (Gadus morhua). Canadian Journal of Fisheries and Aquatic Sciences 54: 104–112.

Lambert, Yvan, and Jean-Denis Dutil. 2000. Energetic consequences of reproduction in Atlantic cod (Gadus morhua) in relation to spawning level of somatic energy reserves. Canadian Journal of Fisheries and Aquatic Sciences 57: 815-825.

LeCren, Eric D. 1951. The length-weight relationship and seasonal cycle in gonad weight and condition in the perch (Perca fluviatilis). Journal of Animal Ecology 20: 201-219.

Louisiana Department of Wildlife and Fisheries press release. 22 March, 2011. http://www.wlf.louisiana.gov/news/33877 . Accessed 18 January, 2012.

Matlock, Gary C. 1990. The life history of red drum. In Chamberlain, George W., Russell J. Miget, and Michael G. Haby (eds.). Red Drum Aquaculture, 1-21. Texas A&M Sea Grant College Program No. TAMU-SG-90-603. College Station, Texas, United States of America.

Overstreet, Robin M. 1983. Aspects of the biology of the red drum, *Sciaenops ocellatus*, in Mississippi. Gulf Research Reports Supplement 1: 45-68.

Pearson, John C. 1929. Natural history and conservation of redfish and other commercial sciaenids on the Texas Coast. Bulletin of the Bureau of Fisheries 129-214.

Peters, Kevin M., and Robert H. McMichael Jr. 1987. Early life history of red drum, S*ciaenops ocellatus* (pisces: sciaenidae), in Tampa Bay, Florida. Estuaries and Coasts 10(2): 92-107.

Roth, Agatha-Marie F., and Donald M. Baltz, 2009. Short-Term Effects of an Oil Spill on Marsh-Edge Fishes and Decapod Crustaceans, Estuaries and Coasts 32(3): 565–572.

Scharf, Frederick S., and Kurtis K. Schlight. 2000. Feeding habits of red drum (Sciaenops ocellatus) in Galveston Bay, Texas: Seasonal diet variation and predator-prey size relationships. Estuaries and Coasts 23(1): 129-139.

Stevenson, Robert D., and William A. Woods, Jr. 2006. Condition indices for conservation: new uses for evolving tools. Integrative and Comparative Biology 46(6): 1169-1190.

Summers, Kevin L., Leroy Folmar, and Miriam Rodón-Naveira. 1997. Development and testing of bioindicators for monitoring the condition of estuarine ecosystems. Environmental Monitoring and Assessment 47:275-301.

Sutter, Frederick C., Richard S. Waller, and Thomas D. McIlwain. 1986. Species Profile: Life histories and environmental requirements of coastal fishes and invertebrates (Gulf of Mexico): Black drum. US Fish and Wildlife Service, Biological Report 82(11.51). Defense Technology Information Center ADA181585.

Sutton, Stephen G., Tammo P. Bult, and Richard L. Haedrich. 2000. Relationships among fat weight, body weight, water weight, and condition factors in wild Atlantic salmon parr. Transactions of the American Fisheries Society 129: 527-538.